\documentclass[a4paper,11pt]{article}
\pdfoutput=1
\usepackage{jcappub,xcolor}
\usepackage[T1]{fontenc}
\usepackage[shortlabels]{enumitem}

\begin{document}
\title{Negative cosmological constant in the dark energy sector: tests from JWST photometric and spectroscopic observations of high-redshift galaxies}

\author[a]{Nicola Menci}
\author[b]{Shahnawaz A. Adil,}
\author[c]{Upala Mukhopadhyay,}
\author[c]{Anjan A. Sen,}
\author[d,e]{and Sunny Vagnozzi}
\affiliation[a]{Istituto Nazionale di Astrofisica (INAF), Osservatorio Astronomico di Roma, Via Frascati 33, 00078 Monte Porzio Catone (RM), Italy}
\affiliation[b]{Department of Physics, Jamia Millia Islamia, New Delhi-110025, India}
\affiliation[c]{Centre for Theoretical Physics, Jamia Millia Islamia, New Delhi-110025, India}
\affiliation[d]{Department of Physics, University of Trento, Via Sommarive 14, 38122 Povo (TN), Italy}
\affiliation[e]{Istituto Nazionale di Fisica Nucleare (INFN), Trento Institute for Fundamental Physics and Applications (TIFPA), Via Sommarive 14, 38122 Povo (TN), Italy}

\emailAdd{nicola.menci@inaf.it}
\emailAdd{shahnawaz188483@st.jmi.ac.in}
\emailAdd{rs.umukhopadhyay@jmi.ac.in}
\emailAdd{aasen@jmi.ac.in}
\emailAdd{sunny.vagnozzi@unitn.it}

\emailAdd{sunny.vagnozzi@unitn.it}

\abstract{Early observations with the James Webb Space Telescope (JWST) have revealed the existence of an unexpectedly large abundance of extremely massive galaxies at redshifts $z \gtrsim 5$: these are in tension with the predictions not only of the standard $\Lambda$CDM cosmology, but also with those of a wide class of dynamical dark energy (DE) models, and are generally in better agreement with models characterized by a phantom behaviour. Here we consider a model, inspired by string theory and the ubiquity of anti-de Sitter vacua therein, featuring an evolving DE component with positive energy density on top of a negative cosmological constant, argued in an earlier exploratory analysis to potentially be able to explain the JWST observations. We perform a robust comparison of this model against JWST data, considering both photometric observations from the CEERS program, and spectroscopic observations from the FRESCO survey. We show that the model is able to accommodate the JWST observations, with a consistency probability of up to $98\%$, even in the presence of an evolving component with a quintessence-like behaviour (easier to accommodate theoretically compared to phantom DE), while remaining consistent with standard low-redshift probes. Our results showcase the potential of measurements of high-redshift galaxy abundances in tests of fundamental physics, and their complementarity with standard cosmological probes.}
\maketitle

\section{Introduction}
\label{sc:introduction}

The concordance $\Lambda$CDM cosmological model constitutes at present one of the simplest frameworks to describe the evolution of our Universe, and owes its success to its ability to describe a wide range of cosmological and astrophysical observations. One of the cornerstones of current cosmology is dark energy (DE)~\cite{Peebles:2002gy,Frieman:2008sn,Nojiri:2017ncd}, a component with negative pressure and positive energy density driving the late-time acceleration of the Universe, whose discovery in 1998 led to one of the biggest paradigm shifts in physics~\cite{SupernovaSearchTeam:1998fmf,SupernovaCosmologyProject:1998vns}: within $\Lambda$CDM, DE takes the form of a \textit{positive} cosmological constant (CC) $\Lambda$~\cite{Carroll:2000fy}. Over the past two and a half decades, the nature of DE has to a large extent been investigated by means of three classes of cosmological observations: low-redshift ($z \lesssim 2$) probes based on Type Ia Supernovae (SNeIa) and Baryon Acoustic Oscillations (BAO), the latter observed in the clustering of various tracers of the large-scale structure, such as galaxies; and high-redshift ($z \sim 1100$) probes based on the Cosmic Microwave Background (CMB): see Ref.~\cite{Huterer:2017buf} for a recent review on observational probes of DE.~\footnote{Nevertheless, it is worth emphasizing that the sensitivity of the CMB to (late-time) DE is indirectly mostly a low-redshift (geometrical) one, through the effect of DE on the distance to the CMB itself (see e.g.\ the recent discussion in Ref.~\cite{Escamilla:2023oce}), whereas the sensitivity of the CMB to DE perturbations and the dynamics thereof is quite limited, except possibly in the case of models which deviate substantially from the positive CC $\Lambda$, which are however ruled out by observations.} By virtue of the significant efforts devoted to the study of observations, underlying theory, and systematics associated to these probes, CMB, BAO, and SNeIa are generally referred to as standard cosmological probes and enjoy a (rightly deserved) privileged status for what concerns cosmological tests of the expansion of the Universe and of fundamental physics, with particular regard to the nature of DE. However, a series of growing cosmological tensions which appear to challenge the $\Lambda$CDM model~\cite{Perivolaropoulos:2021jda}, and the fact that some of these probes are close to being systematics-limited, are making it imperative to start looking for new independent probes, whose complementarity and synergy with the standard CMB+BAO+SNeIa can greatly enrich the landscape of methods to study the Universe: in this sense, a number of so-called ``emerging'' cosmological probes are rapidly gaining momentum, as beautifully summarized in the recent review of Ref.~\cite{Moresco:2022phi}.~\footnote{Of course tests of DE are not limited to cosmological scales. Various tests of DE on astrophysical, astronomical, and even local scales, have been considered over the past decades, see Refs.~\cite{Hamilton:2015zga,Burrage:2016bwy,Burrage:2017qrf,Vagnozzi:2021quy,Zhang:2021ygh,Ferlito:2022mok,Brax:2022olf,Benisty:2023ofi,Zhang:2023neo,Benisty:2023qcv,Benisty:2023dkn,Benisty:2023vbz,Paraskevas:2023itu,Benisty:2023clf,Kaneta:2023wdr,OShea:2024jjw} for examples in this sense.}

In this context, the long-awaited recent observations of massive, distant galaxies delivered by the James Webb Space Telescope (JWST)~\cite{Gardner:2006ky} have opened a remarkable window not only onto astrophysics and the process of galaxy formation, but also onto novel cosmological tests of the contents of our Universe and more generally fundamental physics~\cite{Labbe:2022ahb,Xiao:2023ghw,Arrabal:2023ghw}. This is possible because the abundance of galactic haloes of dark matter (DM) mass $M$ at redshifts $z \approx 5-10$ is predicted to be exponentially sensitive to the growth factor of perturbations $D(z)$, while also being a strongly decreasing function of redshift, and one which depends crucially on the (model-dependent) time-redshift relation~\cite{DelPopolo:2006gn}.~\footnote{Distant galaxies are also useful as a test of cosmological models through their ages, which of course cannot exceed the age of the Universe: see for instance Refs.~\cite{Jimenez:2019onw,Valcin:2021jcg,Bernal:2021yli,Vagnozzi:2021tjv,Wei:2022plg,Cimatti:2023gil,Costa:2023cmu} for recent works exploring the important of stellar and galactic ages as a test of cosmological models.} On the other hand, if we denote by $\Omega_b$ and $\Omega_m$ the cosmic density parameters for baryons and the total matter component respectively (and with $f_b \equiv \Omega_b/\Omega_m$), the stellar mass of a galaxy $M_{\star}$ cannot exceed the maximal allowed baryonic content of the host halo $f_bM$. This implies that comparing the theoretically \textit{predicted} cumulative comoving stellar number density of massive ($M_{\star} \approx 10^{10}-10^{11}\,M_{\odot}$) galaxies, $n(>M_{\star},z)$, with the \textit{observed} abundance of galaxies of given stellar mass at redshifts $z \approx 5-10$, can potentially provide strong constraints on the expansion history and growth of cosmological perturbations, and therefore also on the nature of DE, in a way which is completely independent of the complex baryonic physics involved in the formation of galaxies. Needless to say, such constraints would be extremely valuable in light of their high complementarity to state-of-the-art determinations from CMB, BAO, and SNeIa, both in terms of range of redshifts/cosmic times probed, as well as potential systematics involved.
	
As hinted to earlier, the possibility of testing the expansion history of the Universe at such intermediate redshifts is particularly valuable in view of the tension between local and high-redshift determinations of the Hubble constant $H_0$ (see e.g.\ Refs.~\cite{Verde:2019ivm,DiValentino:2020zio,DiValentino:2021izs,Abdalla:2022yfr,Schoneberg:2021qvd,Shah:2021onj,Kamionkowski:2022pkx,Hu:2023jqc,Vagnozzi:2023nrq,Verde:2023lmm} for recent reviews). At a significance of $\gtrsim 5\sigma$, the Hubble tension poses what is probably one of the most serious observational challenges to $\Lambda$CDM, and can potentially lead us to completely rethink the nature of DE.~\footnote{With no claims as to completeness, see Refs.~\cite{Yang:2018euj,Guo:2018ans,SolaPeracaula:2018wwm,DiValentino:2019ffd,DiValentino:2019jae,Zumalacarregui:2020cjh,Gomez-Valent:2020mqn,DiValentino:2020evt,Moreno-Pulido:2020anb,Gao:2021xnk,SolaPeracaula:2021gxi,Petronikolou:2021shp,Alestas:2021luu,Moreno-Pulido:2022phq,Roy:2022fif,Sharma:2022ifr,SolaPeracaula:2022hpd,Moshafi:2022mva,Schiavone:2022wvq,Gao:2022ahg,Bernui:2023byc,Ben-Dayan:2023rgt,deCruzPerez:2023wzd,Zhai:2023yny,SolaPeracaula:2023swx,Basilakos:2023kvk,Petronikolou:2023cwu} for examples of works exploring the implications of the Hubble tension for the nature of DE. The possibility of a more speculative ``early dark energy'' component, operative at very high redshift, is also a particularly interesting possibility in this context (see e.g.\ Refs.~\cite{Poulin:2018cxd,Niedermann:2019olb,Braglia:2020bym,Oikonomou:2020qah,Vagnozzi:2021gjh,Karwal:2021vpk,Benevento:2022cql,Reeves:2022aoi,Poulin:2023lkg,Odintsov:2023cli}).} Thus, probing  the expansion history of the Universe at  intermediate redshifts $z\approx 6-10$ constitutes a key opportunity to study the validity of $\Lambda$CDM model in an independent manner, and in an epoch which is otherwise extremely difficult to access (see also Refs.~\cite{Gomez-Valent:2023uof,Tutusaus:2023cms} for recents work which highlighted the relevance of intermediate-redshift DE dynamics in the context of the Hubble tension).

Among the various aspects of fundamental physics which can be put to test through measurements of the abundance of massive galaxies at very high redshift, of particular interest to the present work are the properties of DE. In particular, as explicitly demonstrated by one of us in Ref.~\cite{Menci:2020ybl} using data from the CANDELS survey, the abundances of massive, high-redshift galaxies can provide key constraints on the DE equation of state (EoS) $w$. The reason is that $w$ controls the large-scale behaviour of DE at both background and perturbation level, thereby dramatically affecting the growth of density perturbations and the formation of cosmic structures. In this context, a number of recent works (see e.g.\ Refs.~\cite{Lovell:2022bhx,Boylan-Kolchin:2022kae,Forconi:2023izg,Wang:2022jvx,Forconi:2023hsj}) have demonstrated how initial JWST imaging data from NIRCam observations of the Cosmic Evolution Early Release Science (CEERS) program, which uncovered evidence of a surprisingly abundant population of very massive galaxies at extremely high redshifts $7 \lesssim z \lesssim 10$, could severely challenge the $\Lambda$CDM scenario. Specializing to the implications for DE, it was argued by one of us in Ref.~\cite{Menci:2022wia} that these observations in fact exclude a significant portion of dynamical DE parameter space (including the point corresponding to $\Lambda$CDM), and favor models whose equation of state takes values $w<-1$ at some point after recombination, i.e.\ \textit{phantom} models~\cite{Caldwell:2003vq}, see also Ref.~\cite{Wang:2023ros} for further discussions.~\footnote{For examples of other works showcasing the enormous potential of the early JWST observations in constraining various aspects of fundamental physics, see Refs.~\cite{Biagetti:2022ode,Haslbauer:2022vnq,Hutsi:2022fzw,Gandolfi:2022bcm,Maio:2022lzg,Yuan:2023bvh,Dayal:2023nwi,Ilie:2023zfv,Jiao:2023wcn,Parashari:2023cui,Hassan:2023asd,Lei:2023mke,Yoshiura:2023xkd,Padmanabhan:2023esp,Lin:2023ewc,Su:2023jno,Forconi:2023izg,Gouttenoire:2023nzr,Guo:2023hyp,Huang:2023chx,Bird:2023pkr,Wang:2023ros,Libanore:2023oxf,Wang:2023gla,vanPutten:2023ths}. Taking a step aside from new fundamental physics, a very important possibility of course is that the JWST results may call for a better understanding of galaxy formation, as discussed in a number of recent works~\cite{Ferrara:2022dqw,Qin:2023rtf,Pallottini:2023yqg,Wang:2023xmm,Pacucci:2023oci,Iocco:2024rez}.} Although such results are still affected by uncertainties related to the assumed initial mass function (IMF) of the stellar populations (see e.g.\ Ref.~\cite{Steinhardt:2022ghw}) and to the broadband selection criteria (especially for double-break sources, see for example, see Ref.~\cite{Desprez:2023pif}), it is worth highlighting that recent spectroscopic observations in the redshift range $5 \lesssim z \lesssim 6$ and $8 \lesssim z \lesssim 10$~\cite{Xiao:2023ghw,Arrabal:2023ghw} from the JWST First Reionization Epoch Spectroscopically Complete Observations (FRESCO) and CEERS surveys corroborate the above conclusions, while also placing them on a much stronger footing.

The implications of the early JWST results for phantom DE are particularly interesting in view of the Hubble tension. It is now well understood that consistency with measurements of the acoustic angular scale at low redshifts from BAO requires that the sound horizon be lowered in the presence of a higher Hubble constant~\cite{Bernal:2016gxb,Addison:2017fdm,Lemos:2018smw,Aylor:2018drw,Schoneberg:2019wmt,Knox:2019rjx,Arendse:2019hev,Efstathiou:2021ocp,Cai:2021weh,Keeley:2022ojz}. This implies that phantom DE, an inherently late-time component which does not affect the sound horizon, cannot on its own completely solve the Hubble tension. Nevertheless, within the limits imposed by low-redshift data, phantom DE can at least partially alleviate the Hubble tension, in light of its ability to accommodate a higher Hubble constant while keeping the distance to the CMB, and thereby the acoustic angular scale observed in the CMB $\theta_s$, fixed (given that the sound horizon is unaffected, see Ref.~\cite{Alestas:2020mvb} for a recent explicit discussion of parameter degeneracies in the presence of phantom DE). For this and other reasons, the possibility of phantom DE playing some role in the context of the Hubble tension is one which has been give serious consideration in the literature (see e.g.\ Refs.~\cite{Zhao:2017cud,Mortsell:2018mfj,Li:2019yem,Vagnozzi:2019ezj,Alestas:2020zol,Yang:2021flj,Kumar:2021eev,Teng:2021cvy,Heisenberg:2022lob,Chudaykin:2022rnl,Sharma:2022oxh,Ballardini:2023mzm,Dahmani:2023bsb,Montani:2023xpd,daCosta:2023mow} for recent discussions on the subject).

Our discussions so far have been data-driven. However, theory considerations have a lot to offer to the discussion on viable DE models. Firstly, the tiny value of the positive CC $\Lambda$ required to explain observations is at severe odds with the value expected from theory considerations, when interpreted in terms of zero-point vacuum energy density of quantum fields: this is the well-known CC problem~\cite{Weinberg:1988cp,Padmanabhan:2002ji}. Moving on to the simplest ``quintessence'' models for DE, based on a single, minimally coupled scalar field in the absence of higher derivative operators, and with canonical kinetic term~\cite{Wetterich:1987fm,Peebles:1987ek,Ratra:1987rm,Wetterich:1994bg,Caldwell:1997ii}, it is well known that these predict $w>-1$, and cannot therefore give rise to phantom DE, which instead requires a violation of the null energy condition~\cite{Vikman:2004dc,Carroll:2004hc,Nojiri:2005sx,Oikonomou:2022wuk,Trivedi:2023zlf}. It is also worth noting that, in the absence of additional ingredients, these ``vanilla'' quintessence models also worsen the Hubble tension, and one could therefore argue that they are observationally disfavored if the Hubble tension is to be taken seriously~\cite{Vagnozzi:2018jhn,OColgain:2018czj,Colgain:2019joh,Banerjee:2020xcn,Lee:2022cyh}.
 
Although most quintessence scalar field models feature a ground state with positive energy density, corresponding to a de Sitter (dS) vacuum, such a scenario has proven extremely difficult to construct in a controlled setting within string theory. In fact, it has been conjectured that string theory may be unable to accommodate dS vacua~\cite{Danielsson:2018ztv}, as advocated by the swampland program~\cite{Vafa:2005ui,Palti:2019pca,Grana:2021zvf}, whose cosmological implications, particularly for inflation and DE, are far-reaching to say the least (see e.g.\ Refs.~\cite{Obied:2018sgi,Agrawal:2018own,Achucarro:2018vey,Garg:2018reu,Kehagias:2018uem,Kinney:2018nny,Ooguri:2018wrx,Odintsov:2020zkl,Oikonomou:2020oex,Cicoli:2021skd}).~\footnote{Two possible counterexamples to these conjectures are the KKLT~\cite{Kachru:2003aw} and Large Volume Compactification~\cite{Balasubramanian:2005zx} scenarios, although there is still no complete consensus on whether the resulting uplifted (meta)stable dS vacua are sufficiently long-lived to be able to agree with observations.} On the other hand, anti-dS (AdS) vacua, which correspond to a negative CC (nCC), appear ubiquitously within string theory, and are among the best understood quantum gravity backgrounds by virtue of the AdS/CFT correspondence~\cite{Maldacena:1997re}. Of course, a nCC $\Lambda<0$ with energy density $\rho_{\Lambda}<0$ is unable, on its own, to give rise to cosmic acceleration. However, an evolving DE component with positive energy density $\rho_x>0$ on top of a nCC can be consistent with the observed late-time acceleration, provided $\rho_x+\rho_{\Lambda}$ is positive around the present time, and amounts to about 70\% of the total energy budget. Such a scenario, which could be interpreted in terms of a quintessence field whose potential features a negative minimum (AdS vacuum), is of great theoretical interest in light of the previous string-driven considerations. In fact, a number of recent works have explored similar scenarios in light of standard CMB, BAO, and SNeIa observations (see e.g.\ Refs.~\cite{Cardenas:2002np,Poulin:2018zxs,Dutta:2018vmq,Visinelli:2019qqu,Ruchika:2020avj,DiValentino:2020naf,Calderon:2020hoc,Sen:2021wld,Malekjani:2023ple,Adil:2023exv}), finding that these models perform equally well as $\Lambda$CDM, or are even potentially statistically preferred.~\footnote{Another related very interesting possibility which has been explored in the literature involves an AdS phase around recombination, as in the case of so-called AdS-early dark energy models~\cite{Ye:2020btb,Ye:2020oix,Ye:2021nej,Jiang:2021bab,Ye:2021iwa,Jiang:2022qlj,Jiang:2022uyg,Jiang:2023bsz,Ye:2023zel,Peng:2023bik,Wang:2024dka}.} The strong theoretical motivation behind such models, as well as their consistency with standard cosmological probes, strongly motivates further studies thereof, and in particular a comparison against other available observations, potentially in different redshift ranges, where the deviations from $\Lambda$CDM can be substantially more pronounced.

In Ref.~\cite{Adil:2023ara}, it was shown by some of us that a DE sector featuring an evolving component on top of a nCC can, within regions of parameter space consistent with standard cosmological observations, drastically alter the growth of structure at high redshifts with respect to $\Lambda$CDM to the point of \textit{potentially} restoring concordance with the (photometric) JWST CEERS observations. We stress the word ``potentially'' as this earlier analysis was merely an exploratory one: in fact, besides only having explored a few benchmark points in parameter space, a careful treatment of corrections required for an accurate comparison to the JWST observations was lacking, as well as a more complete assessment of the complementarity with standard cosmological probes (the last two points were, in fact, both explicitly mentioned as motivation for a more detailed follow-up work in the closing paragraph of Ref.~\cite{Adil:2023ara}). Given the enormous promise shown by a model featuring an evolving DE component on top of a nCC, it is our goal in the present work to revisit the model and go beyond the exploratory analysis of Ref.~\cite{Adil:2023ara}, by providing a full comparison of such a model against the JWST CEERS observations, while also including information from standard cosmological probes. To place our results on a much more solid footing from the observational point of view, we also consider spectroscopic data from the JWST FRESCO survey~\cite{Xiao:2023ghw}. Overall, our work confirms and reinforces the extremely promising conclusions reached earlier in Ref.~\cite{Adil:2023ara}, and provides further motivation for exploring dark sectors featuring components with negative energy densities, while also highlighting once more the enormous potential held by observations of the abundance of massive galaxies at very high redshift in testing fundamental physics.

The rest of this paper is then organized as follows. In Sec.~\ref{sec:ncc} we briefly review the DE models considered in the rest of the work. Our analysis methods, including the adopted datasets, are discussed in Sec.~\ref{sec:methods}. The results of our analysis, and in particular the resulting limits in dynamical DE parameter space, are presented in Sec.~\ref{sec:results}. Finally, in Sec.~\ref{sec:conclusions} we draw concluding remarks and outline a number of potentially interesting avenues for follow-up work.

\section{Dark Energy models}
\label{sec:ncc}

Following the earlier work of Ref.~\cite{Adil:2023ara}, we consider a DE sector consisting of a cosmological constant $\Lambda \gtrless 0$ which in principle can take either sign (with positive or negative signs corresponding to a dS or AdS vacuum respectively), and with associated energy density $\rho_{\Lambda}=\Lambda/8\pi G$ of the same sign. On top of this dS or AdS vacuum we place an evolving DE component with strictly positive energy density $\rho_x(z)>0$. Rather than committing to a specific microphysical model for the evolving DE component, we assume that this is described by a time-evolving EoS $w_x(z)$ of the Chevallier-Polarski-Linder (CPL) form~\cite{Chevallier:2000qy,Linder:2002et}:
\begin{eqnarray}
w_x(z) = w_0 + w_a\frac{z}{1+z}\,.
\label{eq:wxzcpl}
\end{eqnarray}
There are various reasons why we adopt the widely used CPL parametrization, ranging from its highly manageable 2-dimensional nature, to its direct connection to several physical DE models (including several quintessence DE models, see e.g.\ Refs.~\cite{Linder:2002et,Linder:2006sv,Linder:2007wa,Linder:2008pp,Scherrer:2015tra}) and, last but not least, for ease of comparison to the earlier work of Ref.~\cite{Adil:2023ara}.~\footnote{See Ref.~\cite{Colgain:2021pmf} for a recent discussion on potential shortcomings of the CPL parametrization. For completeness, we note that a number of other parametrizations for the EoS of dynamical DE components have been proposed in the literature, see e.g.\ Refs.~\cite{Efstathiou:1999tm,Jassal:2004ej,Gong:2005de,Barboza:2008rh,Ma:2011nc,Pantazis:2016nky,Yang:2017alx,Pan:2017zoh,Yang:2018qmz,Singh:2023ryd}, see also Ref.~\cite{Perkovic:2020mph} for discussions on the theoretical viability of these parametrizations.} We refer to the evolving component being quintessence-like (phantom) at a given epoch if, at the redshift in question, $w(z)>-1$ ($<-1$). While $w_0$ controls the current (quintessence-like or phantom) nature of DE, whether or not this can change in the past is determined by the value of $w_a$, given that at asymptotically early times ($z \to \infty$) the DE EoS tends to the value $w_0+w_a$. Finally, we note that models crossing which can cross between the two regimes are typically referred to as ``quintom'' models, and have been widely studied in the literature~\cite{Feng:2004ff,Guo:2004fq,Cai:2007gs,Cai:2007qw,Zhang:2009un,Saridakis:2009ej,Cai:2012yf,Bahamonde:2018miw,Leon:2018lnd,Panpanich:2019fxq}, see Ref.~\cite{Cai:2009zp} for a review.

If we denote by $\rho_{\rm crit}^{(0)}$ the current critical energy density, by $\rho_x^{(0)}$ the current energy density of the evolving DE component, and by $\Omega_x \equiv \rho_x^{(0)}/\rho_{\rm crit}^{(0)}$ its density parameter, the energy density of the evolving DE component as a function of redshift is given by the following:
\begin{eqnarray}
\rho_x(z)=\Omega_x\rho_{\rm crit}^{(0)}(1+z)^{3(1+w_0+w_a)}\exp \left ( -3w_a \dfrac{z}{1+z} \right ) \,. \nonumber \\
\label{eq:rhoxz}
\end{eqnarray}
We work under the assumption of a spatially flat Friedmann-Lema\^{i}tre-Robertson-Walker Universe filled, besides the evolving DE component described previously, by the cosmological constant $\Lambda \gtrless 0$, alongside the usual matter and radiation fluids (with density parameters $\Omega_m$ and $\Omega_r$ respectively).~\footnote{We neglect neutrinos for simplicity, given their very limited impact at the cosmological epochs of interest.} Under these assumptions, and defining the density parameter of the cosmological constant $\Omega_{\Lambda} \equiv \Lambda/3$, the evolution of the Hubble rate is governed by the following equation:
\begin{eqnarray}
\frac{H^2(z)}{H_0^2} &=& \Omega_r(1+z)^4 + \Omega_m(1+z)^3 + \Omega_{\Lambda} \nonumber \\
&+&\Omega_x(1+z)^{3(1+w_0+w_a)}\exp \left ( -3w_a \dfrac{z}{1+z} \right )\,.
\label{eq:friedmann}
\end{eqnarray}
where the density parameters satisfy $\Omega_r+\Omega_m+\Omega_{\Lambda}+\Omega_x=1$. Finally, as noted in Ref.~\cite{Adil:2023ara}, it is natural to identify the combination of the evolving CPL component and the cosmological constant $\Lambda$ as comprising the combined DE sector, whose total density parameter is $\Omega_{\text{DE}} \equiv \Omega_x+\Omega_{\Lambda}$. The important thing to note is that, although $\Lambda$ itself and therefore $\Omega_{\Lambda}$ can be negative, the \textit{total} DE density and therefore $\Omega_{\text{DE}}$ have to be positive in order to be able to drive cosmic acceleration and maintain agreement with cosmological observations. Roughly speaking, such a DE sector can in principle be compatible with cosmological observations provided $\Omega_{\text{DE}} \approx 0.7$, which of course is possible even if $\Omega_{\Lambda}<0$, as noted in several recent works~\cite{Dutta:2018vmq,Visinelli:2019qqu,Sen:2021wld,Colgain:2024clf,Favale:2024sdq}. We note that, in order to agree with observations, a more negative $\Omega_{\Lambda}$ needs to be compensated by more negative values of $w_x$, moving towards the phantom regime. These considerations lead to comparatively weak, order unity upper limits on $\vert \Omega_{\Lambda} \vert$, as discussed for instance in Refs.~\cite{Visinelli:2019qqu,Sen:2021wld}, whereas different combinations of $w_x$, $\Omega_x$, and $\Omega_{\Lambda}$ (the latter two summing to the same value of $\Omega_{\text{DE}}$) can lead to a very rich phenomenology, discussed in detail in Ref.~\cite{Calderon:2020hoc} (see also Ref.~\cite{Dash:2023scq} for discussions on the expected sensitivity to a nCC from future 21-cm observations).

From a theoretical standpoint, such a phenomenological model has been argued to loosely carry string motivation in the case where $\Lambda<0$~\cite{Visinelli:2019qqu,Sen:2021wld,Adil:2023ara}. While AdS vacua appear ubiquitously in string theory, the evolving DE component on top can, broadly speaking, be justified on the grounds that string compactifications typically predict the existence of a plethora of ultralight (pseudo)scalar particles. This usually goes under the name of ``string axiverse'' (see for example Refs.~\cite{Svrcek:2006yi,Arvanitaki:2009fg,Cicoli:2012sz,Visinelli:2018utg,Cicoli:2023opf}), with the axion-like particles arising from Kaluza-Klein reduction of higher-dimensional form fields on the topological cycles of the compactification space. The topology of the compactification manifold fixes the number of particles, typically of order hundreds or more, and with masses spread over a huge number of decades, with the rough expectation that the distribution of the logarithms of the masses should be approximately uniform~\cite{Kamionkowski:2014zda,Emami:2016mrt,Karwal:2016vyq}, see also Ref.~\cite{Marsh:2011gr,Ruchika:2020avj} for related studies. The important thing to note is that the effective EoS of multiple interacting scalar fields can be phantom~\cite{Carroll:2003st,Vikman:2004dc,Carroll:2004hc,Deffayet:2010qz,Sawicki:2012pz}: this motivates the possibility of a component with positive energy density, but with EoS $w_x(z)$ potentially crossing $-1$ [as in Eq.~(\ref{eq:wxzcpl}) for suitable choices of $w_0$, $w_a$], sitting on top of a nCC.

Concerning the AdS vacuum itself, the exploratory results of Ref.~\cite{Adil:2023ara} which our work seeks to confirm show that the relevant region of parameter space is one where $\vert \Omega_{\Lambda} \vert \sim {\cal O}(1)$, i.e.\ where the nCC is of the same order of the dS vacuum energy in $\Lambda$CDM (see also Refs.~\cite{Dutta:2018vmq,Visinelli:2019qqu,Sen:2021wld}). Therefore, the magnitude of the nCC is expected to be $\lesssim 10^{-123}$ in Planck units, and one could legitimately worry that this would introduce a (negative) cosmological constant problem. Intriguingly, recent work in string theory~\cite{Demirtas:2021nlu,Demirtas:2021ote} has led to the explicit construction of supersymmetric AdS$_4$ vacua of the right magnitude. This has been achieved within the context of type IIB string theory in compactifications on orientifolds of Calabi-Yau threefold hypersurfaces, with the resulting solution preserving ${\cal N}=1$ supersymmetry, and with the key point of the construction being the perfect cancellation of all perturbative terms in the superpotential. Such a construction explicitly shows that it is possible to obtain an exponentially small nCC within string compactifications, thereby providing further string motivation for the region of parameter space we shall explore in this work.

Before moving on we note that, once the combination of evolving CPL component and $\Lambda$ is identified as making up the combined DE sector, a natural quantity characterizing the latter is the effective equation of state $w_{\text{eff}}(z)$, which can be determined through the twice contracted Bianchi identity, and is given by (see Ref.~\cite{Adil:2023ara} for the full calculation):
\begin{equation}
w_{\rm eff}(z) = \frac{\Omega_x(1+z)^{2+3w_0+3w_a} \left [ w_0+(w_0+w_a)z \right ] \exp \left ( -3w_a \dfrac{z}{1+z} \right ) -\Omega_{\Lambda}}{\Omega_x(1+z)^{3(1+w_0+w_a)}\exp \left ( -3w_a \frac{z}{1+z} \right ) +\Omega_{\Lambda}}\,, \nonumber \\
\label{eq:weffz}
\end{equation}
for which in general $w_{\rm eff}(z=0) \neq w_0$ unless $\Omega_{\Lambda}=0$. It is also worth noting that if the values of $\Omega_x$, $\Omega_{\Lambda}$, $w_0$, and $w_a$ are such that the total DE energy density goes through zero and therefore switches sign at a certain redshift, the associated effective EoS given by Eq.~(\ref{eq:weffz}) necessarily goes through a pole at the same redshift. This has been discussed in detail in Refs.~\cite{Ozulker:2022slu,Adil:2023ara}, and does not in itself signal a pathology given that $w_{\rm eff}(z)$ is not associated to the dynamics of a single microscopical degree of freedom -- however, this does highlight the importance of focusing on the total DE energy density rather than the effective EoS.

\section{Methods}
\label{sec:methods}

For what concerns the properties of high-redshift galaxies, and more generally the formation of structures, the predictions of our model are controlled by seven parameters: the matter density parameter $\Omega_m$, the baryon density parameter $\Omega_b$ (as it controls the maximal baryonic, and therefore stellar, content of a given host DM halo), the density parameter of the (positive or negative) cosmological constant $\Omega_{\Lambda}$, the Hubble constant $H_0$, the present-day linear theory amplitude of matter fluctuations averaged in spheres of radius $8\,h^{-1}{\text{Mpc}}$ $\sigma_8$, and finally the parameters characterizing the evolving DE component with positive energy density, $w_0$ and $w_a$. We note that the radiation density parameter $\Omega_r$ is essentially fixed by extremely high-precision measurements of the CMB temperature monopole (and is in any case negligible), from which it follows that the density parameter of the evolving DE component $\Omega_x>0$ is fixed by the closure relation $\Omega_r+\Omega_m+\Omega_{\Lambda}+\Omega_x=1$ (which basically reduces to $\Omega_{\text{DE}}=\Omega_x+\Omega_{\Lambda} \approx 1-\Omega_m$), and cannot therefore be treated as a free parameter.

For each set of cosmological parameters discussed above, following Ref.~\cite{Linder:2005in} and as reported in Refs.~\cite{Menci:2020ybl,Menci:2022wia,Adil:2023ara}, we compute the evolution of the matter density contrast $\delta$, from which we obtain the linear growth factor of density perturbations $D(z)$. We note that the evolution of $D(z)$ can differ significantly from that within $\Lambda$CDM. The reason is that the equation for the evolution of $\delta$ (see e.g.\ Ref.~\cite{Adil:2023ara}) depends on both the normalized expansion rate $E(z) \equiv H(z)/H_0$ and its first derivative: a different background expansion therefore directly impacts the growth of structure, which within the model at hand can be either suppressed or enhanced depending on the choice of cosmological parameters, with the value of $\Omega_{\Lambda}$ playing an important role (see Ref.~\cite{Adil:2023ara} for more detailed discussions on this point).

The next step is to compute the predicted maximal abundance of galaxies with stellar mass $M_{\star}$. More precisely, the relevant quantities are the comoving cumulative number or mass  (stellar mass or DM halo mass) densities, which quantify the number or mass density of halos above a given threshold, or the number or stellar mass density of stars contained in galaxies more massive than a given threshold. Here we proceed as in Refs.~\cite{Menci:2020ybl,Menci:2022wia} and just recall the basic steps to compute these quantities, while encouraging the reader to consult the above papers for further details (see also see also Eqs.~(3.7--3.10) of Ref.~\cite{Adil:2023ara} for explicit expressions of the four quantities discussed above). For a given set of the seven cosmological parameters we compute the DM halo mass function $dn(M,z)/dM$, which quantifies the number of DM haloes of mass $M$ per unit mass per unit comoving volume in the mass range $[M;M+dM]$ at a given redshift. We do so following the prescription of Ref.~\cite{Sheth:1999mn}, itself an extension of the Press-Schechter formalism~\cite{Press:1973iz} accounting for ellipsoidal collapse.

From the DM halo mass function we can obtain the \textit{maximal} comoving number density of galaxies with stellar mass in the range $[M_{\star};M_{\star}+dM_{\star}]$, $dn_{\star}(M_{\star},z)/dM_{\star}$. This is given by $dn_{\star}(M_{\star},z)/dM_{\star} = f_bdn(M,z)/dM$, where $f_b \equiv \Omega_b/\Omega_m$ is the cosmic baryon fraction. We stress that this is the \textit{maximal} density because it is computed under the extremely optimistic assumption that the entire available baryonic reservoir ends up being converted into stars. This assumption is of course unrealistic, as in reality the efficiency of converting gas into stars, typically denoted by $\epsilon$, is of order $\epsilon \lesssim 0.2$ or less, with a moderate redshift dependence~\cite{Leroy:2008kh,Combes:2010vc,Tacchella:2018qny}. However, this assumption is extremely conservative for the purposes of our study: given that galaxies cannot outnumber their DM haloes, one can exclude those cosmological models for which \textit{even when $\epsilon=1$} the predicted number density of galaxies of given stellar mass at a certain redshift falls short of the observed abundance of galaxies within the same range of redshift and stellar mass. To put it differently, a galaxy of stellar mass $M_{\star}$ can only form if a DM halo of mass $M_{\star}/\epsilon f_b$ has formed first. Finally, we can compute the corresponding cumulative comoving maximal number and stellar mass densities of galaxies with stellar masses larger than a given observational threshold $\overline{M_{\star}}$. These are obtained by integrating $dn(M,z)/dM$ and $f_bMdn(M,z)/dM$ respectively, where the lower limit of the integration range is given by the threshold DM halo mass $\overline{M_{\star}}/f_b$,~\footnote{The upper limit of the integration range is formally $\infty$, although the integrand drops exponentially quickly for sufficiently large DM halo mass.} and averaging over the cosmic volume $V(z_1,z_2)$ enclosed between the redshift range $z_{\min}-z_{\max}$ covered by the observations (the average is performed within the integral, again see Eqs.~(3.7--3.10) of Ref.~\cite{Adil:2023ara} for explicit expressions of these quantities). Two further corrections are required before the computed cumulative comoving stellar mass density of galaxies, $\rho_{\star}(>\overline{M_{\star}},z)$, can be compared against the JWST observations, as we will discuss shortly.

Our goal is now to compute $\rho_{\star}(>\overline{M_{\star}},z)$ as a function of various choices of cosmological parameters, while ensuring that the latter are in agreement with standard cosmological probes. This will allow us to discuss the complementarity between the latter, and the observed abundance of high-redshift galaxies from JWST. Operationally, we therefore proceed as follows. The cosmological datasets we consider are given below:~\footnote{While more recent datasets are available for some of the measurements below (especially for what concerns BAO and SNeIa data), the reason why we chose these datasets was for consistency with the work of Ref.~\cite{Sen:2021wld}, in order to make use of the chains which were produced therein. Nevertheless, we expect very minimal quantitative changes were we to use the most up-to-date eBOSS BAO data~\cite{eBOSS:2020yzd} and \textit{PantheonPlus} SNeIa data~\cite{Scolnic:2021amr}, and no qualitative changes, given that our conclusions are almost entirely driven by the JWST observations. We therefore expect them to be robust against the use of slightly more update BAO and SNeIa data.}
\begin{itemize}
\item \textit{CMB data}: measurements of CMB temperature anisotropy and polarization power spectra, their cross-spectra, and CMB lensing power spectrum reconstructed from the temperature 4-point correlation function, from the \textit{Planck} 2018 legacy release. This combination is usually referred to as \texttt{TTTEEE}+\texttt{lowl}+\texttt{lowE}+\texttt{lensing}, and we analyse it making use of the official \textit{Planck} likelihood~\cite{Planck:2018vyg,Planck:2018lbu,Planck:2019nip}.
\item \textit{BAO data}: isotropic and anisotropic distance and expansion rate measurements from the SDSS-MGS, 6dFGS, and BOSS DR12 collaborations~\citep{Ross:2014qpa,Beutler:2011hx,BOSS:2016wmc}.
\item \textit{SNeIa data}: distance moduli measurements from the \textit{Pantheon} SNeIa sample within the redshift range $0.01<z<2.3$~\cite{Pan-STARRS1:2017jku}.
\item \textit{SH0ES prior}: a prior on the Hubble constant $H_0=(73.30 \pm 1.04)\,{\text{km}}/{\text{s}}/{\text{Mpc}}$, as determined by the SH0ES team~\cite{Riess:2021jrx}.
\end{itemize}
Considering the combination of the above datasets, we derive constraints on our set of cosmological parameters by making use of Markov Chain Monte Carlo (MCMC) methods, with predictions for the cosmological observables in question derived using the publicly available Boltzmann solver \texttt{CLASS}~\cite{Lesgourgues:2011re,Blas:2011rf}. We make use of the publicly available cosmological MCMC sampler \texttt{MontePython 3.3}~\cite{Audren:2012wb,Brinckmann:2018cvx}. We monitor the convergence of the generated MCMC chains through the Gelman-Rubin parameter $R-1$~\cite{Gelman:1992zz}, requiring $R-1<0.02$ for our chains to be considered converged, and analyze them using the \texttt{GetDist} package~\cite{Lewis:2019xzd}. For a full discussion of the methodology and the resulting constraints, we refer the reader to Ref.~\cite{Sen:2021wld}.

In the next stage of the analysis, we introduce the JWST observations, in particular the inferred/observed comoving cumulative stellar mass density $\rho_{\text{obs}}(>\overline{M_{\star}})$. More specifically, we consider the value of $\rho_{\text{obs}}(>\overline{M_{\star}})$ inferred from two different classes of observations:
\begin{itemize}
\item the six most massive, intrinsically red galaxies in the redshift range $9 \lesssim z \lesssim 11$, identified in the first NIRCam observations of the JWST CEERS program, as reported in Ref.~\cite{Labbe:2022ahb}, and which we treat as a measurement of $\rho_{\text{obs}}(>\overline{M_{\star}})$ at $z_{\text{eff}}=10$;
\item the three most massive, optically dark (dust-obscured) galaxies with robust spectroscopic redshifts, in the redshift range $5 \lesssim z \lesssim 6$, identified within the JWST FRESCO NIRCam/grism survey, as reported in Ref.~\cite{Xiao:2023ghw}, and which we treat as a measurement of $\rho_{\text{obs}}(>\overline{M_{\star}})$ at $z_{\text{eff}}=5.5$.
\end{itemize}
From the two above observations the inferred stellar mass density is of the order of $\gtrsim 10^6\,M_{\odot}{\text{Mpc}}^{-3}$ and $\gtrsim 10^5\,M_{\odot}{\text{Mpc}}^{-3}$ respectively.

From the MCMC chains described above, we then select only those points which are consistent within $2\sigma$ with the standard cosmological probes discussed earlier. We stress that in our MCMC analysis we have allowed for $\Omega_{\Lambda} \gtrless 0$, i.e.\ either a dS or AdS vacuum energy component. Each of the models consistent within $2\sigma$ with the standard cosmological observations is then compared against the inferred values of $\rho_{\text{obs}}(>\overline{M_{\star}})$ from the JWST CEERS and FRESCO observations described above. However, in order to properly carry out this comparison, we need to account (and correct) for the fact that both the values of $\rho_{\text{obs}}(>\overline{M_{\star}})$ reported by Ref.~\cite{Labbe:2022ahb} and Ref.~\cite{Xiao:2023ghw} have been obtained assuming a certain fiducial cosmology, in this case $\Lambda$CDM with a given choice of parameters. Let us refer to the vector of fiducial cosmology parameters as $\boldsymbol{f}$, whereas we denote by $\boldsymbol{\theta}$ the vector of cosmological parameters for each point in our MCMC chain. As discussed in detail in Ref.~\cite{Menci:2022wia}, one then needs to apply two different corrections:
\begin{itemize}
\item the inferred $\rho_{\text{obs}}(>\overline{M_{\star}})$ has to be corrected by a ``volume factor'' $f_{\text{vol}} \equiv V_{\boldsymbol{f}}/V_{\boldsymbol{\theta}}$ to appropriately rescale the volume density, where $V$ is the cosmic volume computed for the specific model and choice of cosmological parameters in question, at the effective redshift $z_{\text{eff}}$;
\item similarly, the measured masses have to be corrected by a ``luminosity factor'' $f_{\text{lum}} \equiv d_{L,\boldsymbol{f}}/d_{L,\boldsymbol{\theta}}$ to account for the fact that the stellar masses have been inferred from the observed luminosities within the assumption of the given fiducial cosmology, where $d_L$ is the luminosity distance computed for the specific model and choice of cosmological parameters in question, once more at the effective redshift $z_{\text{eff}}$.
\end{itemize}
The cosmology-corrected maximal comoving cumulative stellar mass densities can then be directly compared to the JWST observations described above.

We consider a given cosmological model (which, we recall, is already ensured to be consistent within $2\sigma$ with the standard cosmological probes) to be excluded if the maximal predicted stellar mass density of galaxies at the given effective redshift is lower than the value inferred from JWST CEERS or FRESCO observations. From this we can then compute the exclusion probability $P$ for a given model, or conversely the probability of consistency with the JWST observations $Q$ (obviously $P+Q=1$). This is performed using the procedure described in Ref.~\cite{Menci:2022wia}, through a Monte Carlo procedure which also accounts for uncertainties in the observational estimates of masses -- in particular potential systematics related to the spectral energy distribution (SED) fitting procedure -- and statistical errors in the observational number densities.

More in detail, for each set of cosmological parameters which are consistent with the standard cosmological probes after running our MCMC, we simulate the dispersion of the observed number density $n(M_{\star})$ due to both systematics affecting $M_{\star}$ and statistical errors affecting $n(M_{\star})$. For the former, we conservatively assign an uncertainty of $0.5\,{\text{dex}}$ to the stellar masses $M_{\star}$ reported by CEERS. This uncertainty budget accounts rather generously for systematic uncertainties related to the SED fitting procedure (see e.g.\ Refs.~\cite{Santini:2014ghw,Menci:2022wia}), particularly for what concerns the ages of stellar populations, dust extinction, metallicity, and assumed shapes of the star formation histories. When simulating the uncertainties associated to the CEERS values of $M_*$ in our Monte Carlo procedure, the limit stellar mass $\overline{M_*}$ is extracted from such an interval following an uniform distribution to simulate systematic uncertainties. For what concerns the spectroscopic FRESCO observations, the uncertainties in the estimated stellar masses are simulated in our Monte Carlo procedure by extracting an uncertainty $\Delta M_*$ following a Gaussian distribution with variance given by the uncertainties reported in Ref.~\cite{Xiao:2023ghw}. The observed number densities of galaxies associated to stellar masses $M_{\star}$ extracted as described above are compared to the theoretical expectations for the number density of DM haloes with mass $M=M_*/f_b$, and we estimate upper and lower Poisson confidence limits using the statistical method presented in Ref.~\cite{Ebeling:2003tf}, which presents improved numerical approximations to assess the statistical significance (Poissonian confidence limits) of an observed small number of events, which is precisely the case at play here. This allows us to derive a confidence level $P$ for the probability that the observed number densities are below the theoretical predictions. Notice that the the value $f_b$ is not taken to be a free parameter, but is computed from the ratio $f_b\equiv \Omega_b/\Omega_m$\ where $\Omega_b$ and $\Omega_m$ are taken from the chains of cosmological parameters discussed above.

\section{Results}
\label{sec:results}

We begin by comparing our theoretical predictions against the stellar mass density inferred from the most massive objects identified by NIRCam observations of the JWST CEERS program, as reported in Ref.~\cite{Labbe:2022ahb}. In particular, these observations refer to the redshift range $z_{\min}=9 \lesssim z \lesssim z_{\max}=11$, which we treat at an effective redshift $z_{\text{eff}}=10$, focusing on the most massive bin considered in Ref.~\cite{Labbe:2022ahb}, corresponding to $M_{\star} \geq  \overline{M_{\star}}=10^{10.5}M_{\odot}$. As discussed in Sec.~\ref{sec:methods}, from the MCMC chains we extract cosmological parameter vectors consistent within $2\sigma$ with the standard CMB, BAO, and SNeIa cosmological probes, predict the corresponding maximal comoving cumulative stellar mass density, which we then correct for the assumed fiducial cosmology, and then compare this against the JWST CEERS observations. For each model, we derive the exclusion probability $P$, or conversely the probability of consistency with the JWST observations, $Q=1-P$.

\begin{figure}[!ht]
\centering
\includegraphics[width=0.8\linewidth]{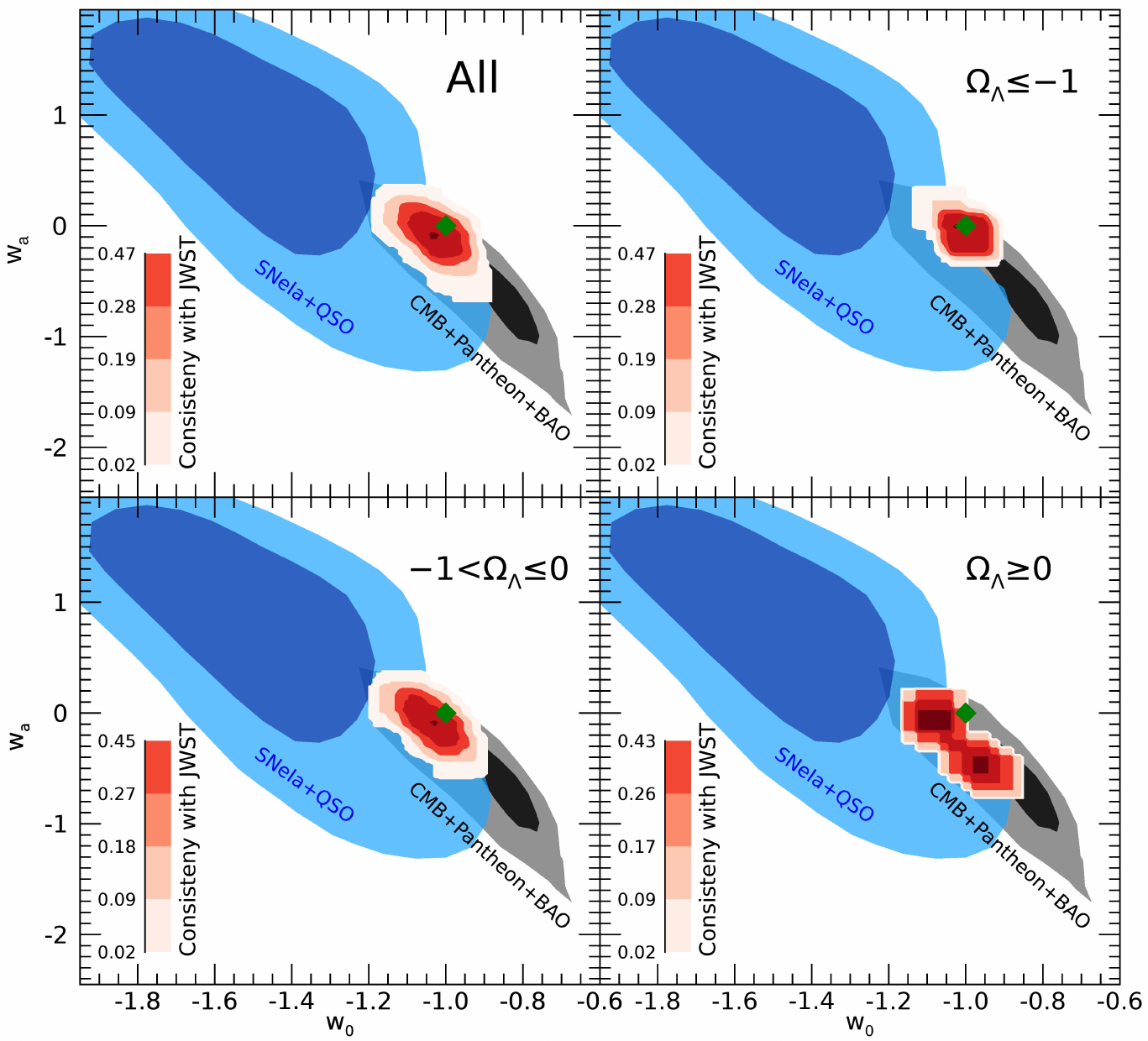}
\caption{Contours in the $w_0$-$w_a$ plane colored by the probability of consistency with the stellar mass density inferred from the six most massive galaxies identified in the first NIRCam observations of the JWST CEERS program, in the redshift range $9 \lesssim z \lesssim 11$, as reported in Ref.~\cite{Labbe:2022ahb}. The contours are reported considering the dynamical DE model discussed in Sec.~\ref{sec:ncc}, featuring an evolving DE component with positive energy density, sitting on top of a vacuum energy component with density parameter $\Omega_{\Lambda} \gtrless 0$. The four different quadrants correspond to different regimes for the value of $\Omega_{\Lambda}$: all values $\Omega_{\Lambda} \gtrless 0$ (upper left quadrant), $\Omega_{\Lambda}<-1$ (upper right quadrant), $-1<\Omega_{\Lambda}<0$ (lower left quadrant), and $\Omega_{\Lambda}>0$ (lower right quadrant). As discussed in Sec.~\ref{sec:methods}, we only consider combinations of cosmological parameters consistent within $2\sigma$ with the standard CMB+BAO+SNeIa cosmological probes. Our contours are compared to the $2\sigma$ and $3\sigma$ contours allowed by the latest CMB+BAO+SNeIa measurements (black contours) as reported in Ref.~\cite{Brout:2022vxf}, and the combined Hubble diagram of SNeIa and quasars (QSOs, blue contours) reported in Ref.~\cite{Risaliti:2018reu}. The green dot corresponds to the $\Lambda$CDM case ($w_0=-1$, $w_a=0$).}
\label{fig:contours_labbe}
\end{figure}

The results of our analysis are reported in terms of consistency probability contour plots as a function of the evolving DE parameters $w_0$ and $w_a$ in Fig.~\ref{fig:contours_labbe}, where each of the four quadrants refers to different regimes for the vacuum energy density parameter $\Omega_{\Lambda}$: dS vacuum $\Omega_{\Lambda}>0$ (lower right quadrant), AdS vacuum with $-1< \Omega_{\Lambda} < 0$ (lower left quadrant), AdS vacuum with even more negative $\Omega_{\Lambda}<-1$ (upper right panel), and finally any value of $\Omega_{\Lambda}$ (upper left quadrant). Superimposed on the same plots are $2\sigma$ and $3\sigma$ joint confidence intervals in the $w_0$-$w_a$ plane obtained from a combination of CMB, BAO, and SNeIa measurements (black contours, as reported in Ref.~\cite{Brout:2022vxf}), as well as from the combined Hubble diagram of SNeIa and quasars (QSOs, blue contours, as reported in Ref.~\cite{Risaliti:2018reu}), reaching up to redshift $z \sim 5.5$.

The first noteworthy result is that the existence of a region of $w_0$-$w_a$ parameter space featuring a DE component with an AdS vacuum which is consistent with all the standard cosmological probes \textit{and} the abundance of high-redshift galaxies inferred from JWST. This can be contrasted to the earlier results of Ref.~\cite{Menci:2022wia} which only considered a dS vacuum, finding that a major portion of the $w_0$-$w_a$ parameter space favored by the standard cosmological probes is excluded at significance $>2\sigma$ by the JWST observations. This result on its own, therefore, represents a solid quantitative confirmation of the earlier exploratory results of Ref.~\cite{Adil:2023ara}, and confirms that the presence of a nCC in the DE sector can help accelerate structure formation, thereby aiding the formation of very massive objects at very high redshift beyond what is possible within $\Lambda$CDM (for more detailed explanations of why this occurs, we refer the reader to Ref.~\cite{Adil:2023ara}).

It is also interesting to note that the favored region in $w_0$-$w_a$ parameter space is not far from ($w_0$, $w_a$)=($-1$, $0$), while lying slightly within the phantom regime, in qualitative agreement with Ref.~\cite{Adil:2023ara}. Within this region, we are able to achieve a consistency probability with the JWST measurements of up to $47\%$. Moreover, this same region is also consistent within $\approx 2\sigma$ with the region favored by SNeIa+QSOs. As $\Omega_{\Lambda}$ is increased, and we therefore move from the upper right quadrant to the two lower quadrants, we see a gradual shift in the required properties of the evolving DE component: this progressively shifts from being characterized by mostly quintessence-like behaviour (potentially crossing the phantom divide in the past, in agreement with the results of Ref.~\cite{Adil:2023ara}), to a quintessence-like behaviour at present but a phantom behaviour in the past, i.e.\ moving towards the lower right part of the $w_0$-$w_a$ plane. The latter point is in excellent agreement with the earlier findings of Ref.~\cite{Menci:2022wia}, which indeed focused precisely on this regime, and showed that therein most of the $w_a>0$ region is excluded, unless $w_0$ is deep in the phantom regime. Either way, it is clear that as $\Omega_{\Lambda}$ is increased, the point ($w_0$, $w_a$)=($-1$, $0$) is progressively disfavored, with a probability of consistency of $<10\%$ in the $\Omega_{\Lambda}>0$ case (lower right quadrant), in agreement with Ref.~\cite{Menci:2022wia}. In the latter case, one should pay attention to the fact that Fig.~2 in Ref.~\cite{Menci:2022wia} plots the \textit{exclusion} probability, whereas here we are plotting the probability for \textit{consistency}: when one accounts for this, it is easy to see that the results are consistent with those of Ref.~\cite{Menci:2022wia}. On the other hand, an AdS vacuum plus evolving DE component scenario displays a good consistency with the JWST measurements (see especially the upper right quadrant): the reason is that the presence of the nCC helps accelerate structure formation, thereby reducing the need for phantom DE, which also happens to be theoretically much harder to accommodate with respect to a quintessence-like component.

While the above conclusions are robust on the theoretical side, critical issues may affect the measurements of Ref.~\cite{Labbe:2022ahb} we are comparing against. In first place, potential uncertainties may affect the calibration of the JWST photometric data used therein. A second issue may concern the Chabrier IMF adopted by Ref.~\cite{Labbe:2022ahb} to derive stellar masses. While we do not expect that assuming other universal forms for the IMF based on low-redshift conditions would change (or even strengthen) the constraints we have derived, the star formation process can be significantly different at redshifts as high as the ones we are concerned with (see e.g.\ Ref.~\cite{Steinhardt:2023ghw}).

For such reasons, it is important to perform an independent test against spectroscopic data in a different redshift range. With this in mind, we then proceed to carry out a comparison against the stellar mass density inferred from the three most massive, optically dark (dust-obscured) galaxies with robust spectroscopic redshifts from the JWST FRESCO survey~\cite{Xiao:2023ghw}. Note that here as well the stellar masses have been conservatively inferred using a Chabrier IMF. We expect that assuming other universal forms for the IMF (e.g.\ Salpeter, Kennicut) would yield similar or even larger values of $M_{\star}$, and therefore our results can be interpreted as being conservative in this sense. As in our previous comparison, we conservatively consider for the measured value $M_{\star}$ the one corresponding to the lower tip of the uncertainty provided by Ref.~\cite{Xiao:2023ghw}. As for the uncertainties in the number density, the confidence levels thereon are derived using the statistical method presented in Ref.~\cite{Ebeling:2003tf}, as discussed in Ref.~\cite{Menci:2022wia}.

\begin{figure}[!ht]
\centering
\includegraphics[width=0.8\linewidth]{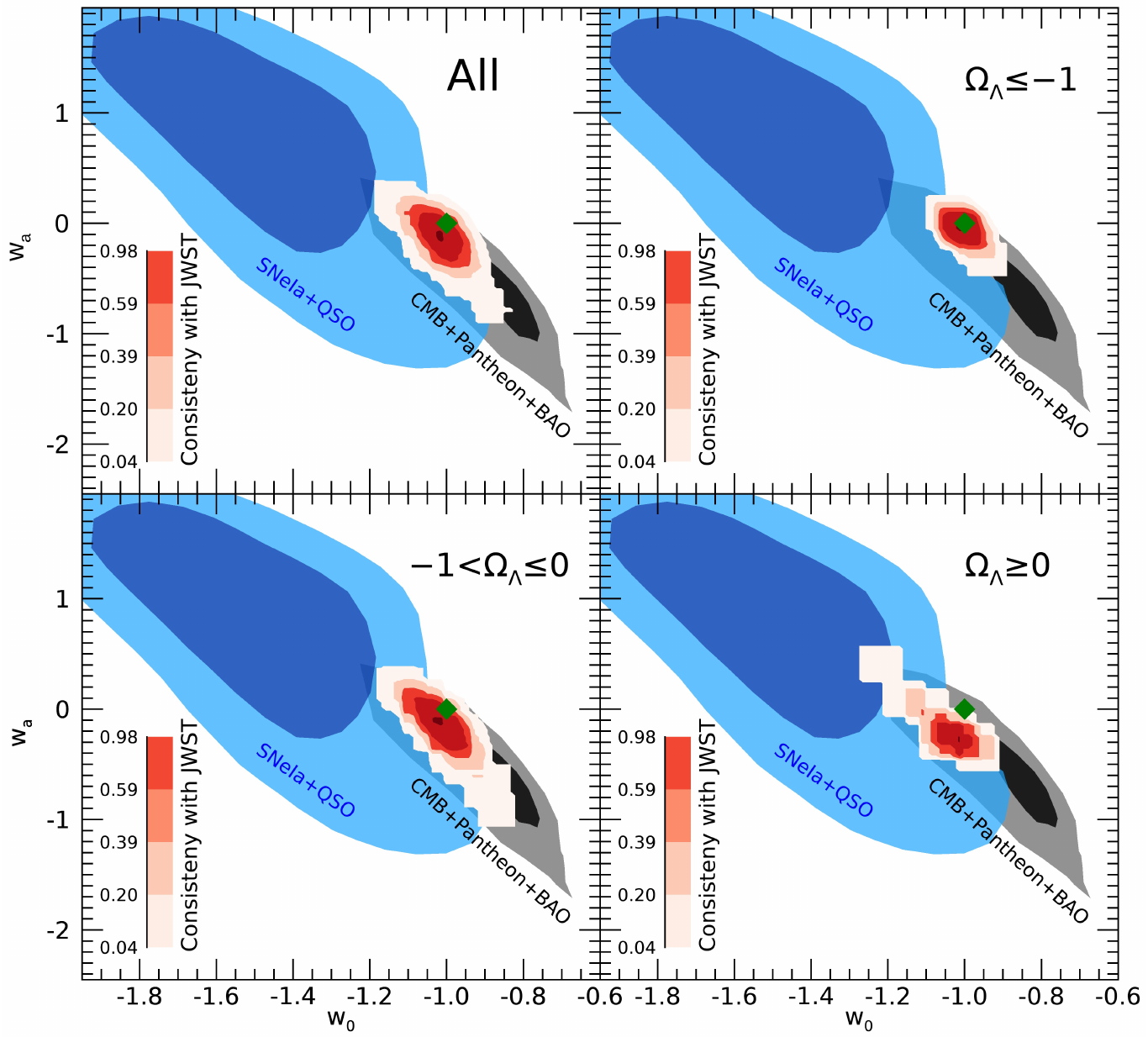}

\caption{Same as in Fig.~\ref{fig:contours_labbe}, but considering the three most massive, optically dark (dust-obscured) galaxies with robust spectroscopic redshifts identified within the JWST FRESCO survey, in the redshift range $5 \lesssim z \lesssim 6$, as reported in Ref.~\cite{Xiao:2023ghw}.}
\label{fig:contours_xiao}
\end{figure}

The results of such a comparison are shown in Fig.~\ref{fig:contours_xiao}, with the same quadrant structure as in Fig.~\ref{fig:contours_labbe}. We see that the results are completely consistent, and very similar, to those obtained comparing against the photometric sample at higher redshift of Ref.~\cite{Labbe:2022ahb}. The only quantitative difference is that overall we are able to achieve higher levels of consistency with the JWST FRESCO data, as high as $98\%$. Moreover, just as observed earlier, we see that the fraction of evolving DE models with phantom behaviour increases with increasing values of $\Omega_{\Lambda}$, confirming the trend seen with the photometric sample: in other words, introducing a nCC helps accelerate structure formation and reduce the need for a phantom DE component, a welcome reduction from the theory point of view.

We stress that the CEERS and FRESCO datasets adopted are not the only JWST-based measurements of the abundance of high-redshift galaxies. The reason why we have adopted these and not others, however, is that at present they constitute the most constraining datasets, and are therefore best suited for the scope of the present work. For instance, NIRCam observations of the GLASS-ERS 1324 program reported in Ref.~\cite{Santini:2023ghw} involve galaxies at comparable redshifts but with much lower stellar masses (log $M_{\star}/M_{\odot} \lesssim 9.5$), and are therefore not expected to significantly tighten existing cosmological constraints. On the other hand, JWST observations of the Hubble Ultra Deep Field (HUDF) and UKIDSS Ultra Deep Survey field reported in Ref.~\cite{Navarro-Carrera:2023ytd} extend to lower redshifts ($z\lesssim 8$) compared to ones we considered. Indeed, the measurements reported in Ref.~\cite{Navarro-Carrera:2023ytd} yield number densities which are comparable to, or even lower than, those reported in Ref.~\cite{Grazian:2015ghw} for the CANDELS/UDS, GOODS-South, and HUDF fields. The cosmological implications of the measurements reported in Ref.~\cite{Grazian:2015ghw} were analyzed by one of us in Ref.~\cite{Menci:2020ybl}, finding much weaker constraints on dynamical DE compared to the ones obtained in Ref.~\cite{Menci:2022wia} based on JWST CEERS data, and in the present work. Therefore, we do not expect that the measurements presented in Refs.~\cite{Santini:2023ghw,Navarro-Carrera:2023ytd} would add much constraining power to constraints on dark energy from standard cosmological probes, although a study thereof could be interesting for follow-up works. The CEERS and FRESCO measurements we consider in the present work provide at present an unprecedented combination of high redshifts, large masses, and large abundances which make them uniquely suited for our study.

\begin{figure}[!ht]
\centering
\includegraphics[width=0.8\linewidth]{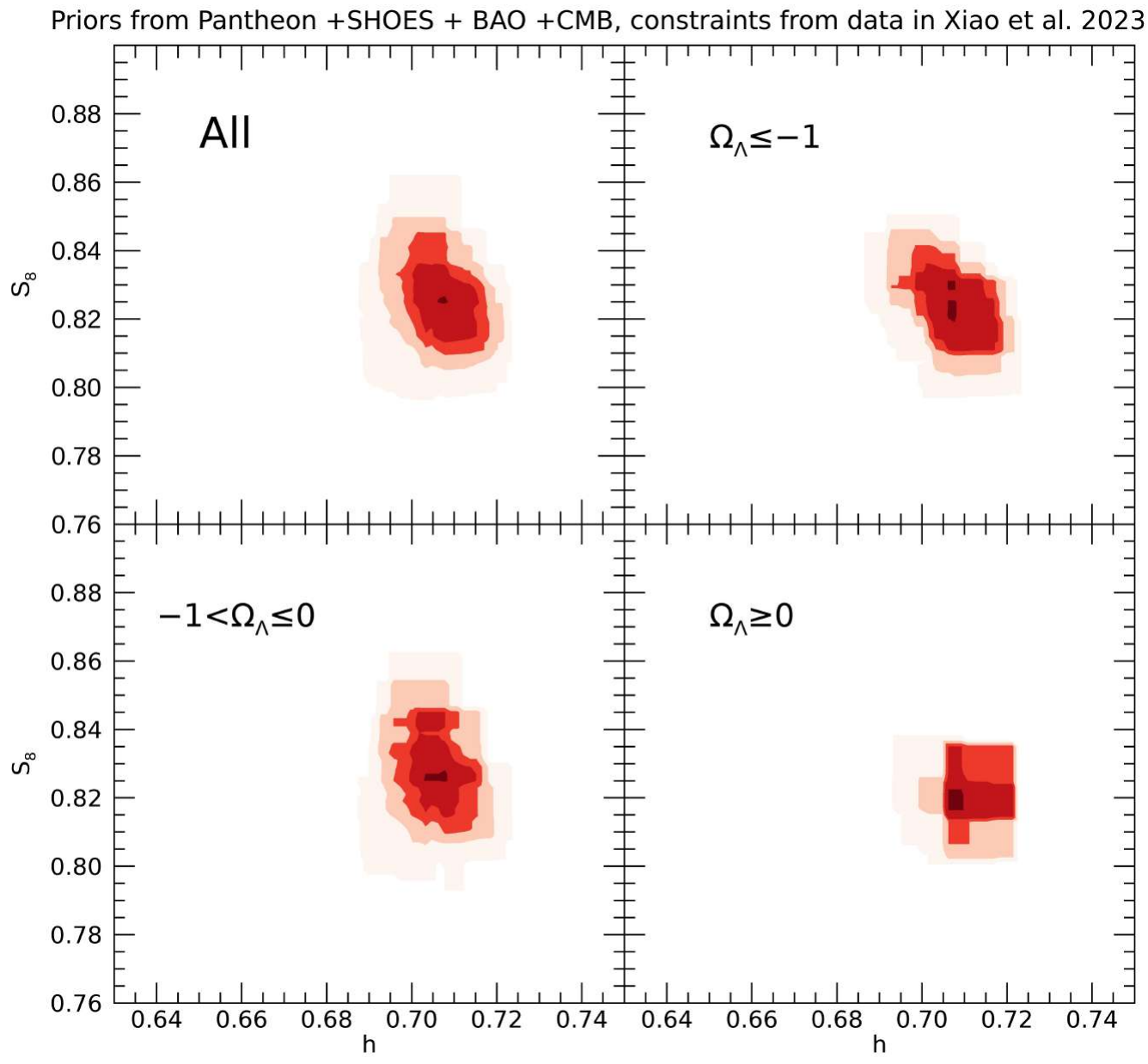}
\caption{Contours of probability of consistency with the same observations considered in Fig.~\ref{fig:contours_xiao}, and with the same color coding, in the plane of the reduced Hubble constant $h \equiv H_0/(100\,{\text{km}}/{\text{s}}/{\text{Mpc}})$ and the clustering parameter $S_8\equiv \sigma_8\sqrt{\Omega_m/0.3}$.}
\label{fig:s8h0}
\end{figure}

For completeness, in Fig.~\ref{fig:s8h0} we display the associated contours in the $h$-$S_8\equiv \sigma_8\sqrt{\Omega_m/0.3}$ parameter space, where $h \equiv H_0/(100\,{\text{km}}/{\text{s}}/{\text{Mpc}})$ is the reduced Hubble constant, and $S_8$ controls the overall strength of matter clustering in the late Universe. While the contours we obtain are, by construction, consistent with the standard cosmological observables we considered, they display some tension with weak lensing (and, to some extent, galaxy clustering) measurements, which favour lower values of $S_8$ (see e.g.\ Refs.~\cite{DiValentino:2018gcu,DiValentino:2020vvd,Nunes:2021ipq}). This is of course not unexpected.

Before closing, a comment is in order regarding our treatment of the cosmic baryon fraction $f_b$. As explained earlier, we have not treated $f_b$ as a free parameter, but as a derived parameter which is self-consistently computed from the values of $\omega_b$, $\Omega_m$, and $H_0$ at each point in the MCMC chains. This approach differs from a number of other studies on the cosmological implications of the JWST measurements, where $f_b$ is either fixed to the best-fit value inferred within $\Lambda$CDM from the \textit{Planck} 2018 measurements, or freely varied within some prior range. We believe our approach is somewhat more robust as it ensures that $f_b$ is self-consistently constrained by the other probes at play, varying consistently with the same parameters which control our predictions for the growth of structure and thereby the stellar mass density. Of course, it is worth noting that the inferred values of $f_b$ are model-dependent, and can be affected by the choice of both pre- and post-recombination models. Indeed, it is known that the assumed pre-recombination model can affect the inferred value of $\Omega_mh^2$, a fact which has been appreciated in the context of the Hubble tension~\cite{Jedamzik:2020zmd}, whereas $\Omega_bh^2$ is essentially fixed by considerations on Big Bang Nucleosynthesis and the relative height of the odd versus even acoustic peaks in the CMB. We note that higher (lower) values of $f_b$ would lead to proportionally higher (lower) values of the cumulative comoving number and stellar mass densities of galaxies (as both scale linearly with $f_b$), thereby slightly weakening (strengthening) our existing constraints. Nevertheless, within the range of allowed model-dependent values of $f_b$, the effect is expected to be small, and changes to the growth history associated to a different dark energy model are expected to be much larger.

\section{Conclusions}
\label{sec:conclusions}

The puzzling abundance of extremely massive galaxies at very high redshift unveiled by the early JWST observations has the potential to upturn the current concordance $\Lambda$CDM model, itself already plagued by other observational tensions. At the same time, the JWST observations can be used to test alternative cosmological models, and can potentially rule out those models which do not allow for a sufficiently fast growth of structure required to explain the formation of the galaxies observed by JWST. As argued by various earlier works~\cite{Menci:2020ybl,Menci:2022wia,Adil:2023ara}, observations of the abundance of high-redshift galaxies can place strong constraints on dynamical DE models, in a way which is highly complementary to standard CMB, BAO, and SNeIa cosmological probes at low redshifts. In the earlier exploratory work of Ref.~\cite{Adil:2023ara} some of us argued that a DE sector featuring a negative cosmological constant (AdS vacuum) with an evolving DE component on top, which carries strong motivation from string theory given the ubiquity of AdS vacua therein, can lead to a more efficient growth of structure at early times, potentially explaining the JWST observations. Our goal in this work has been that of going beyond this exploratory analysis, by \textit{a)} performing a more careful comparison of such a model against JWST data, while also \textit{b)} better assessing the complementarity with standard cosmological probes, and \textit{c)} considering also spectroscopic data from the JWST FRESCO survey~\cite{Xiao:2023ghw} in order to place the results on a more solid footing.

Our results, which qualitatively confirm the exploratory findings of Ref.~\cite{Adil:2023ara}, can be summarized as follows:
\begin{itemize}
\item a DE sector featuring an evolving component with positive energy density on top of a negative cosmological constant with density parameter $\Omega_{\Lambda}<0$ can indeed improve consistency with photometric JWST observations at redshifts $9 \lesssim z \lesssim 11$ (in agreement with the findings of Ref.~\cite{Adil:2023ara}), with consistency of up to $47\%$;
\item as $\Omega_{\Lambda}$ is increased, the behaviour evolving DE component progressively shifts towards a more phantom behaviour -- to put it differently, introducing a negative cosmological constant allows for better agreement with the JWST observations with a decreased need for phantom behaviour, which is more problematic compared to a quintessence-like behaviour from a theoretical point of view;
\item for $\Omega_{\Lambda}>0$, the JWST observations favor dynamical DE models which cross the phantom divide, in agreement with the findings of Ref.~\cite{Menci:2022wia};
\item these findings remain intact when considering spectroscopic observations from the JWST FRESCO survey at $5 \lesssim z \lesssim 6$, placing all the earlier conclusions on a much more robust footing from the observational point of view;
\item the JWST observations, and more generally observations of the abundance of high-redshift galaxies, are highly complementary to standard cosmological probes from CMB, BAO, and SNeIa, as well as observations of high-redshift QSOs.
\end{itemize}
In short, we have shown that DE sectors featuring a negative cosmological constant are extremely interesting from the perspective of the high-redshift galaxies observed by JWST. Moreover, this class of models, and more generally models featuring negative energy densities in the DE sector, also have the potential to partially (albeit not completely) alleviate the Hubble tension (see e.g.\ Refs.~\cite{Visinelli:2019qqu,Akarsu:2019hmw,DiValentino:2020naf,Acquaviva:2021jov,Akarsu:2021fol,Akarsu:2021max,Sen:2021wld,Akarsu:2022typ,Adil:2023exv,Akarsu:2023mfb}). This therefore confirms that such models are interesting from all three the observational, phenomenological, and also theoretical perspective, in the latter case in light of their strong motivation from string theory.

Overall, our results further strengthen the case for illuminating the nature of DE, and more generally testing new fundamental physics, using observations of high-redshift galaxies, in a redshift range which cannot be reached by standard cosmological probes. We believe such observations have the potential to become an important ``emergent probe'' (see e.g.\ Refs.~\cite{Moresco:2022phi}) in coming years, especially with the achievable higher spatial resolution and improved sensitivity from future ALMA/NOEMA and deep JWST spectroscopic observations, whose complementarity with upcoming CMB measurements~\cite{SimonsObservatory:2018koc,SimonsObservatory:2019qwx} it would be interesting to explore. From the theory side, it could be interesting to extend our analysis to one implementing a fundamental string scenario from first principles, along the lines of Refs.~\cite{Cicoli:2018kdo,Ruchika:2020avj,Oikonomou:2023kqi}). The early JWST observations have the potential to shake modern cosmology and shed new light onto the nature of the dark sector components. Our work represents a small step in the latter direction, and we cannot wait to see what lies in store for cosmology as our long-awaited space telescope keeps gathering data.

\section*{Acknowledgments}
N.M. acknowledges support from Ministero dell'Universit\`{a} e della Ricerca (MUR, Italian Ministry for Universities and Research) through the Progetti di Rilevante Interesse Nazionale (PRIN) project ``Black Hole winds and the Baryon Life Cycle of Galaxies: the stone-guest at the galaxy evolution supper'' (grant agreement no.\ 2017-PH3WAT), and INAF Theory Grant ''AGN-driven outflows in cosmological models of galaxy formation''. A.A.S. and U.M. acknowledge support from the Science and Engineering Research Board (SERB) of the Government of India through research grant no.~CRG/2020/004347. S.A.A., U.M., and A.A.S. acknowledge the use of the High Performance Computing facility Pegasus at IUCAA, Pune, India. S.V. acknowledges support from the University of Trento and the Provincia Autonoma di Trento (PAT, Autonomous Province of Trento) through the UniTrento Internal Call for Research 2023 grant ``Searching for Dark Energy off the beaten track'' (DARKTRACK, grant agreement no.\ E63C22000500003), and from the Istituto Nazionale di Fisica Nucleare (INFN) through the Commissione Scientifica Nazionale 4 (CSN4) Iniziativa Specifica ``Quantum Fields in Gravity, Cosmology and Black Holes'' (FLAG). This publication is based upon work from the COST Action CA21136 ``Addressing observational tensions in cosmology with systematics and fundamental physics'' (CosmoVerse), supported by COST (European Cooperation in Science and Technology).

\bibliographystyle{JHEP}
\bibliography{nccjwstjcap}

\end{document}